\documentclass[prd,twocolumn,showpacs,superscriptaddress,usenatbib]{revtex4}

\def\plottwo#1#2{\centering \leavevmode
\epsfxsize=.45\textwidth \epsfbox{#1} \hfil
\epsfxsize=.45\textwidth \epsfbox{#2}}

\usepackage{epsfig}
\usepackage{graphicx}
\usepackage{amsmath,amssymb}
\usepackage{aas_macros}     
\newcommand{\cmnt}[1]{{}}
\begin{document}
%
%
\title{Molecular hydrogen in the cosmic recombination epoch}
%
\author{Esfandiar Alizadeh}
\email{ealizad2@illinois.edu}
\affiliation{Department of Physics, University of Illinois at Urbana-Champaign, 1110 West Green Street, Urbana, IL 61801}

\author{Christopher M. Hirata}
\email{chirata@tapir.caltech.edu}
\affiliation{Caltech M/C 350-17, Pasadena, CA 91125, USA}

\date{December 9, 2010}
%
%
\begin{abstract}
The advent of precise measurements of the cosmic microwave background (CMB) anisotropies has motivated correspondingly precise calculations of the cosmic recombination history.  Cosmic recombination proceeds far out of equilibrium because of a ``bottleneck'' at the $n=2$ level of hydrogen: atoms can only reach the ground state via slow processes: two-photon decay or Lyman-$\alpha$ resonance escape.  However, even a small primordial abundance of molecules could have a large effect on the interline opacity in the recombination epoch and lead to an additional route for hydrogen recombination.  Therefore, this paper computes the abundance of the H$_2$ molecule during the cosmic recombination epoch. Hydrogen molecules in the ground electronic levels  X$^1\Sigma^+_g$ can either form from the excited H$_2$ electronic levels B$^1\Sigma^+_u$ and C$^1\Pi_u$ or through the charged particles H$_2^+$, HeH$^+$ and H$^-$. We follow the transitions among all of these species, resolving the rotational and vibrational sub-levels. Since the energies of the X$^1\Sigma^+_g$--B$^1\Sigma^+_u$ (Lyman band) and X$^1\Sigma^+_g$--C$^1\Pi_u$ (Werner band) transitions are near the Lyman-$\alpha$ energy, the distortion of the CMB spectrum caused by escaped H~Lyman-line photons accelerates both the formation and the destruction of H$_2$ due to this channel relative to the thermal rates.  This causes the populations of H$_2$ molecules in X$^1\Sigma^+_g$ energy levels to deviate from their thermal equilibrium abundances.   We find that the resulting H$_2$ abundance is $10^{-17}$ at $z=1200$ and $10^{-13}$ at $z=800$, which is too small to have any significant influence on the recombination history.     
\end{abstract}

\maketitle
%
\section{introduction}
The era of percent-level precision cosmology started with the exquisite measurements of the cosmic microwave background (CMB) anisotropies by the {\slshape Wilkinson Microwave Anisotropy Probe} ({\slshape WMAP}) satellite \cite{2003ApJS..148....1B}. The CMB anisotropies are a very useful tool for cosmologists for two reasons. First, the shapes and normalizations of the temperature and polarization spectra are sensitive to a host of cosmological parameters \cite{1996PhRvD..54.1332J}. Second, the physics underlying the CMB power spectrum is thought to be well understood.  It can be calculated by linear perturbation theory of the Einstein and Boltzmann equations around a homogeneous, isotropic background \cite{1970ApJ...162..815P,1980PhRvD..22.1882B,1983ApJ...274..443B}; the perturbation equations can be solved rapidly by modern numerical codes that have achieved agreement at the 0.1\%\ level in code comparisons \cite{2003PhRvD..68h3507S}.  However, to solve these equations one needs to know the number density of free electrons as a function of redshift $n_e(z)$, the so called recombination history, which enters into these equations through the Thompson scattering of photons from free electrons.

The first cosmological recombination calculations were carried out more than 40 years ago \cite{1968ApJ...153....1P, 1968ZhETF..55..278Z}, showing the importance of non-equilibrium hydrogen recombination because of the high optical depth of the Lyman series lines in the early universe. A hydrogen atom can only reach its ground state from the H{\sc\,i} 2s level or 2p levels via two-photon decay and redshifting out of the Lyman-$\alpha$ line, respectively.  The early analyses assumed Boltzmann equilibrium of all $n\ge 2$ levels of hydrogen, and thus had to follow only ionized hydrogen H$^++e^-$, excited hydrogen H$^\ast(n\ge2)$, and ground-state hydrogen H(1s).

To obtain $n_e(z)$ to high accuracy, it is necessary to include additional physics.  Thus theorists have considered helium recombination \cite{1969PThPh..42..219M, 1971PThPh..46..416M, 1995PhRvD..52.5498H,2005AstL...31..359D, 2007MNRAS.378L..39K, 2008PhRvD..77h3006S, 2008PhRvD..77h3007H, 2008PhRvD..77h3008S, 2008A&A...485..377R, 2010MNRAS.402.1221C}; deviations from Boltzmann equilibrium for the $n\ge 2$ levels of hydrogen \cite{2007MNRAS.374.1310C, 2010PhRvD..81h3005G, 2010MNRAS.407..599C, 2010MNRAS.407..658D, 2010PhRvD..82f3521A}; a host of two-photon processes \cite{2005AstL...31..359D, 2006A&A...446...39C, 2006AstL...32..795K, 2007MNRAS.375.1441W, 2008PhRvD..77h3007H, 2008A&A...480..629C, 2008PhRvD..78b3001H, 2010A&A...512A..53C}; the transport of photons near Lyman-$\alpha$ due to multiple resonant scattering \cite{1989ApJ...338..594K, 1990ApJ...353...21K, 1991Ap.....34..124G, 1994ApJ...427..603R, 2008AstL...34..439G, 2009A&A...496..619C, 2009PhRvD..80b3001H, 2009A&A...503..345C, 2010A&A...512A..53C}; and cross-talk among various lines and the photoionization continuum \cite{2007A&A...475..109C, 2008PhRvD..77h3006S, 2010PhRvD..81h3004K}.

The workhorse recombination code {\sc Recfast}, used for {\slshape WMAP} parameter constraints, was a fitting function to such non-equilibrium calculations including all of the physical processes recognized as important in the year $\sim$2000 \cite{1999ApJ...523L...1S,2000ApJS..128..407S}, and there have been some subsequent updates \cite{2008MNRAS.386.1023W}.  {\sc Recfast} was sufficiently accurate for the observations of its time, however to fully take advantage of the power of the {\slshape Planck} satellite data (launched 2009) it is important to find $n_e(z)$ to the sub-percent level \cite{2006MNRAS.373..561L,2008MNRAS.386.1023W}. This realization triggered a flurry of papers considering a host of new phenomena that could affect the recombination history to the percent and sub-percent level, culminating in two new publicly available codes that properly treat the radiative transfer effects in hydrogen and helium recombination \cite{2010arXiv1010.3631C, 2010arXiv1011.3758A}.

The current paper is a continuation of the same effort to reach the required level of accuracy.  We consider how the formation and destruction of hydrogen molecules (H$_2$) in the X$^1\Sigma_g^+$, B$^1\Sigma_u^+$ and C$^1\Pi_u$ electronic states can change the recombination history.  The reason that H$_2$ might be able to change the recombination history is that the Lyman and Werner bands (X$^1\Sigma_g^+$--B$^1\Sigma_u^+$ and X$^1\Sigma_g^+$--C$^1\Pi_u$) are near the Lyman-$\alpha$ energy ($h\nu_{{\rm Ly}\alpha}=10.2\,$eV).  Thus the excitation, de-excitation, photodissociation, and photoassociation of the H$_2$ molecule can shuffle photons between the red and the blue sides of the Lyman-$\alpha$ line.  In an expanding Universe, a photon redder than Lyman-$\alpha$ is likely to simply redshift and eventually become a part of the far-infrared background, whereas a photon bluer than Lyman-$\alpha$ will redshift into the Lyman-$\alpha$ frequency and excite a ground-state hydrogen atom (which at $z>900$ would have been likely to be photoionized).  At an order of magnitude level, one would expect this H$_2$-mediated redistribution to become possibly significant if the net optical depth in the Lyman and Werner bands $\tau_{\rm LW}$ were of order $10^{-3}$, which for the recombination-era density and Hubble rate, and total oscillator strengths of order unity, would require an abundance $x[{\rm H}_2]\equiv n({\rm H}_2)/n_{\rm H} \sim 10^{-12}$.  Therefore we are interested in even tiny quantities of molecular hydrogen.  It is worth noting that some models of early Universe chemistry have found $x[$H$_2]\gtrsim 10^{-12}$ \cite{1998A&A...335..403G, 2008A&A...490..521S} during the recombination epoch with simplified (i.e. not level-resolved) reaction networks.

The calculation of the abundance of hydrogen molecules has already been considered by many authors {\em but for a different cosmological goal}, that is to assess the effect of the H$_2$ molecule on the cooling of metal-free gas and its implications for primordial star formation \cite{1967Natur.216..976S, 1968ApJ...154..891P,1969PThPh..41..835H, 1972PASJ...24...87Y, 1976ApJ...205..103H, 1984ApJ...280..465L, 1996ApJ...464..523H, 1996ApJ...467..522H, 1997ApJ...474....1T}.  Since direct radiative association to the X$^1\Sigma_g^+$ electronic level, i.e. 2H$\rightarrow$H$_2($X$^1\Sigma_g^+)+\gamma$ is forbidden, two separate hydrogen atoms must reach the ground state of H$_2$ through an intermediate route. For the case of the post-recombination era when $k_{\rm B} T_{\text{CMB}} < 0.2\,$eV the accessible routes are through the H$_2^+$ \cite{1967Natur.216..976S} and H$^-$ \cite{1968ApJ...154..891P,1969PThPh..41..835H} intermediate states.  Indeed, complex reaction networks have been constructed to follow hydrogen chemistry \cite{1984ApJ...280..465L, 1998A&A...335..403G, 1998ApJ...509....1S, 2000MNRAS.316..901F, 2002JPhB...35R..57L, 2006MNRAS.372.1175H}.  These have identified in particular the significance of the recombination-induced CMB spectral distortion \cite{2006A&A...458L..29C, 2008A&A...485..377R, 2008A&A...488..861C} in controlling pregalactic photochemistry, the importance of rate coefficients \cite{2006ApJ...640..553G, 2008MNRAS.388.1627G, 2009MNRAS.393..911G}, and the importance of following transitions among the various rotational and vibrational levels of the H$_2^+$ ion at $z<500$ \cite{2006MNRAS.372.1175H}.

However, at the redshift of interest for this paper, $z \sim 1000$, there is another route for the formation of hydrogen molecules: the inverse Solomon process \cite{2006ApJ...646L..91D}.  At this era there are enough ultraviolet photons (both blackbody photons and spectral distortion photons) to facilitate the photo-attachment of two hydrogen atoms into an excited H$_2$ molecule in one of the rovibrational levels of either the B$^1\Sigma^+_u$ (Lyman band) or C$^1 \Pi_u$ (Werner band) electronic states with energies $\sim 10\,$eV.  The excited H$_2$ molecule will re-emit the photon and decay to either a bound H$_2$(X$^1\Sigma_g^+$) molecule, or to the continuum of the X level (i.e. to two H atoms).  In equation form,
\begin{equation}
2{\rm H} + \gamma \leftrightarrow {\rm H}_2({\rm B}^1\Sigma_u^+,{\rm C}^1\Pi_u) \leftrightarrow {\rm H}_2({\rm X}^1\Sigma_g^+) + \gamma.
\label{eq:H2process}
\end{equation}
This mechanism and the charged-particle processes (H$^-$, H$_2^+$, and HeH$^+$) control the H$_2$ abundance at high redshift.  The possible effect on hydrogen atom recombination is the main focus of this paper.  We note that Ref.~\cite{2006ApJ...646L..91D} found only a small production of H$_2$ via this mechanism, but they did not include the spectral distortion photons in their rate coefficient and hence the {\em total} rate of H$_2$ production could be many orders of magnitude larger.  Of course, the same spectral distortion also drives H$_2$ photodissociation -- the left arrows in Eq.~(\ref{eq:H2process}) -- so the net effect on the H$_2$ abundance requires a detailed calculation.  Deviations from thermal equilibrium abundances arise not from the amplitude of the ultraviolet photon spectrum, but the way in which its peculiar shape beats against the forest of H$_2$ lines and dissociation continua.  Since we work at $z>800$ we will not distinguish the matter versus radiation temperature in this paper.
 
This paper is organized as follows: in Sec.~\ref{sec:AC} we write down the rate equations for the bound-bound and bound-free transitions. These equations are then solved in the steady state approximation and the results are presented in Sec.~\ref{sec:R}. We discuss the size of the H$_2$ abundances found in the previous chapter on the absorption of Ly-$\alpha$ photons and conclude in Sec.~\ref{sec:DandC}.


\section{Abundance calculation}
\label{sec:AC}

In the standard hydrogen recombination calculation \citet{2000ApJS..128..407S} one follows the evolution of several hundred energy levels of the H{\sc\,i}, He{\sc\,i}, and He{\sc\,ii} atoms by including all bound-bound and bound-free transitions and treating the radiative transfer of line photons in the expanding universe using the Sobolev approximation \citep{1957SvA.....1..678S,1994ApJ...427..603R}.  A similar treatment can be used for the H$_2$ molecule.

This section is organized as follows: we begin with a description of the reactions included (Sec.~\ref{ss:br}) and then turn to the rate equations (Sec.~\ref{ss:re}) and the steady-state approximation (Sec.~\ref{ss:ss}).  Finally, we describe our model for the molecular data (Sec.~\ref{ss:moldata}) and the radiation field (Sec.~\ref{ss:rf}), which is required in order to evaluate the transition rates among H$_2$ levels.  The inclusion of the charged-particle reactions is described in Sec.~\ref{ss:charged}.
 
 \subsection{Basic reactions}
 \label{ss:br}
 
Here, we consider the ground X and the excited B and C electronic states of the hydrogen molecule.  The latter are technically not bound since they can undergo spontaneous radiative dissociation, but they are long-lived. We designate them with their rotational ($J$) and vibrational ($\nu$) quantum numbers; in the case of the C levels, which are $\Lambda$-doubled, it is necessary to describe the parity as either vector-like [C$^+$, $P=(-1)^J$] or axial [C$^-$, $P=-(-1)^J$].  We consider only the bound energy levels of these states up to rotational quantum number of $J=20$. 

The  bound-free radiative reactions involving these levels (we do not consider the collisional reactions in this paper) are the dipole-allowed transitions,
\begin{equation}
{\rm H}_2({\rm B}^1\Sigma^+_u)  \leftrightarrow {\rm H}(1{\rm s})+{\rm H}(1{\rm s})+ \gamma,
\label{BtoCont}
\end{equation}
and
\begin{equation}
{\rm H}_2({\rm C}{^{\pm}}\,{^1}\Pi_u)  \leftrightarrow {\rm H}(1{\rm s})+{\rm H}(1{\rm s}) + \gamma.
\label{CtoCont}
\end{equation}
The bound-bound reactions are the dipole-allowed transitions
\begin{equation}
{\rm H}_2({\rm B}^1\Sigma^+_u) \leftrightarrow {\rm H}_2({\rm X}^1\Sigma^+_g) + \gamma  \label{BtoX} 
\end{equation}
and
\begin{equation}
{\rm H}_2({\rm C}{^{\pm}}\,{^1}\Pi_u) \leftrightarrow {\rm H}_2({\rm X}^1\Sigma^+_g) + \gamma,  \label{CtoX}
\end{equation}
and the quadrupole-allowed transition,
\begin{equation}
{\rm H}_2({\rm X}^1\Sigma^+_g,\nu J) \leftrightarrow {\rm H}_2({\rm X}^1\Sigma^+_g,\nu'J') + \gamma. \label{XtoX}
\end{equation}
There are in principle other quadrupole-allowed transitions; however those involving the B and C electronic states will be small compared to the dipole-allowed transitions (B,C$\rightarrow$X).  We consider quadrupole transitions among the levels of the X electronic state with different rovibrational quantum numbers $\nu J$ because there are no allowed dipole decays from these levels, and hence quadrupole decay might be significant in comparison with excitation by (rare) ultraviolet photons.
 
We will denote the ``thermal abundance'' of an H$_2$ level by its abundance if the reaction 2H$\leftrightarrow$H$_2$ were in equilibrium, i.e.
\begin{eqnarray}
x_{i,{\rm th}} \!\!\! &=& \!\!\! \frac{(2J_i+1)g_eg_{\rm nuc}}{g_{\rm H}^2} \left( \frac{2\pi\hbar^2}{k_{\rm B}T} \frac{m[{\rm H}_2]}{m_{\rm H}^2} \right)^{3/2}
\nonumber \\ && \!\!\! \times
e^{-[E_i - 2E_{{\rm H}(1s)}]/kT} n_{\rm H}x_{{\rm H}(1s)}^2,
\label{eq:thermalabundance}
 \end{eqnarray}
where $g_{\rm nuc}$ is the nuclear degeneracy, $g_e=1$ is the electronic degeneracy, and $g_{\rm H}=4$ is the degeneracy of an H atom.  Note that since the early Universe is not in ionization (Saha) equilibrium, the choice of 2H$\leftrightarrow$H$_2$ as a reference reaction to define the thermal abundance is merely for convenience, and that achieving the thermal abundance does not imply full thermodynamic equilibrium.
 
 \subsection{Rate equations}
 \label{ss:re}
 
 The rate equations for the above reactions can be written as:
 \begin{equation}
 \dot{x}_i=\dot{x}_i \arrowvert_{bb} + \dot{x}_i \arrowvert_{bf}, 
 \label{equ:xdot}
 \end{equation}
 where $i$ denotes the level under consideration (we always resolve rotational and vibrational quantum numbers of H$_2$), the overdot $\dot{}$ denotes a derivative with respect to proper time, and $x_i \equiv n_i/n_{\rm H}$ where $n_{\rm H}$ is the proper density of hydrogen nuclei (in any form -- H{\sc\,ii}, H{\sc\,i}, or H$_2$).
 
The bound-bound term is given by:
\begin{eqnarray}
\dot{x}_i \arrowvert_{bb}
\!\!\! &=& \!\!\!
  - \displaystyle\sum\limits_{j<i}^{} P_{ij}\Bigl\{ x_i A_{ij}[1+f(\nu_{ij+})]-x_j A_{ij}\frac{g_i}{g_j}f(\nu_{ij+}) \Bigr\}  \nonumber \\
&& \!\!\!+ \displaystyle\sum\limits_{j>i}^{} P_{ji}\Bigl\{ x_j A_{ji} [1+f(\nu_{ji+})] -x_i A_{ji} \frac{g_j}{g_i} f(\nu_{ji+}) \Bigr\}.
 \nonumber\\
\label{equ:xdotbb}
\end{eqnarray}
The first sum in the right hand side shows the rate of decrease of $x_i$ by radiative decays to lower levels (spontaneous + stimulated) and increase of $x_i$ by radiative absorption. The second sum is for radiative decays  from the higher levels to the level $i$ and their inverse processes. Here $f(\nu_{ji+}) = f(\nu_{ij}+\epsilon)$ is the photon phase space density on the blue side of the line. Also, $P_{ij}$ is the Sobolev escape probability \cite{1960mes..book.....S} (see Ref.~\cite{2000ApJS..128..407S} for a short derivation) which is the probability that a photon emitted in the line to escape out of it via redshifting before being reabsorbed by a molecule in a lower $j$ level. It is given by:
\begin{equation}
P_{ij}=\frac{1-e^{-\tau_{ij}}}{\tau_{ij}},
\label{tau}
\end{equation}
where the optical depth is 
\begin{equation}
\tau_{ij}=\frac{c^3 n_{\rm H}}{8 \pi H \nu_{ij}^3} A_{ij}\left(\frac{g_i}{g_j}x_j-x_i\right).
\label{equ:Sobolev}
\end{equation}
As a first step, we assume the H$_2$ lines are optically thin, that is $|\tau_{ij}| \ll 1$, and therefore all the emitted photons will escape out of the resonance ($P_{ij} \approx 1$).  This assumption must be checked at the end of the calculation for self-consistency; later we will find it to be extremely good for all lines.

Similarly, the bound-free term can be written as
\begin{eqnarray}
\dot{x}_i \arrowvert_{bf} \!\!\! &=& \!\!\! -x_i \int_{E_{\rm{free}}}^{E_i} \alpha_i(E_f)[1+f(E_i-E_f)] \, dE_f \nonumber\\
 && \!\!\! + n_{\rm H} x_{\rm H(1s)}^2 \int_{E_{\text{free}}}^{E_i}  \beta_i(E_f) f(E_i-E_f) \, dE_f.
\label{equ:xdotbf}
\end{eqnarray}
The first term on the right hand side is for the dissociation of the hydrogen molecule into two H(1s) atoms via radiative decay to an unbound vibrational state, and the second term is for the inverse process.  Here $E_{\text{free}}=-1$ Hartree is the energy of two separated H(1s) atoms with no relative kinetic energy. The functions $\alpha_i(E) dE=A_{i\rightarrow \text{free}}(E) dE$ are the Einstein coefficients for the decay from a bound state $i$ to the continuum with energy $E$. The radiative absorption coefficients $\beta_i$ can then be calculated using the principle of detailed balance:
\begin{equation}
\beta_i(E)=\left( \frac{x_i}{n_H x_H^2} \right)_{\text{th}} \frac{1+f_{\text{th}}(E_i-E)}{f_{\text{th}}(E_i-E)} \alpha_i(E).
\end{equation}
Using equilibrium thermodynamics to find the thermal abundance ratio (Eq.~\ref{eq:thermalabundance}) and plugging in the blackbody spectrum $f_{\text{th}}(E)=1/(e^{E/k_{\rm B} T}-1)$ we finally find
\begin{eqnarray}
\beta_i(E) \!\! &=& \!\!
\frac{(2J_i+1) g_e g_{\rm nuc} }{g_{\rm H}^2}\left(\frac{2 \pi\hbar^2}{k_B T}\frac{m[{\rm H}_2]}{m_{\rm H}^2}\right)^{3/2}
\nonumber \\ && \!\! \times
e^{-(E_i-E_{\text{free}})/k_B T} \alpha_i(E).
\label{equ:beta}
\end{eqnarray}
Here, $g_{\rm H}=4$ is the degeneracy of the ground energy level of the hydrogen atom, $g_e$ is the electron spin degeneracy of the bound state $i$ and is equal to 1 for all the states considered in this paper as they are all in singlet electronic spin states.  Finally, $g_{\rm nuc}$ is the nuclear spin degeneracy of the bound state $i$. It can be calculated by demanding that the total wavefunction change the sign under exchange of the two protons; this implies $g_{\rm nuc}=1$ for the even-$J$ energy levels of the X, B and C$^+$ electronic states and for the odd-$J$ states of C$^{-}$. For the rest of the bound states the protons are in their triplet spin state, i.e $g_{\rm nuc}=3$.    
  
 \subsection{Steady state approximation}
 \label{ss:ss}
 
 We can rewrite the rate equations above in matrix form. To do that it is convenient to define the bound-bound transition matrix: 
\begin{equation}
R_{ij} = \left\{ \begin{array}{lll} A_{ij}[1+f(\nu_{ij})] & & j<i \\
(g_j/g_i)A_{ji} f(\nu_{ij}) & & i<j \\
0 & & i=j. \end{array}\right.
\end{equation}
In addition, we define the dissociation term 
\begin{equation}
\gamma_i=\int_{E_{\rm free}}^{E_i} \alpha_i(E_f)[1+f(E_i-E_f)] \, dE_f,
\label{equ:dissociation}
\end{equation}
and the source term
\begin{equation}
s_i=\int_{E_{\rm free}}^{E_i} n_{\rm H} x_{\rm H(1s)}^2 \beta_i(E_f) f(E_i-E_f)\, dE_f.
\label{equ:source}
\end{equation}
Then it is straightforward to show that the Eqs.~(\ref{equ:xdot}), (\ref{equ:xdotbb}) and (\ref{equ:xdotbf}) can be written in the compact form
\begin{equation}
\dot{x}_i = -\sum_j T_{ij} x_j +s_i
\end{equation}
where the transition matrix ${\bf T}$ is defined by:
\begin{equation}
T_{ij}=\delta_{ij}\left(\sum_{k} R_{ik} + \gamma_i \right) -R_{ji}.
\label{eq:Tdef}
\end{equation}
If the transition times are much smaller than the age of the Universe, i.e. if the smallest eigenvalue $\lambda_{\rm min}$ of ${\bf T}$ is $\gg H$, we can take $\dot{x}_i=0$. Then the abundances can be found by the solution to the linear system:
\begin{equation}
{\bf x} = {\bf T}^{-1}{\bf s}.
\label{equ:final}
\end{equation}
In fact, since the steady-state H$_2$ abundance is exponentially increasing at the end of recombination, we would like $\lambda_{\rm min}$ to be larger than $\zeta H$, where $\zeta = d\ln x_{\rm ss}[$H$_2]/d\ln a$.  The steady-state approximation is found to be valid until $z\approx 810$, i.e. during the portion of the recombination epoch most relevant to CMB anisotropies.  We will find the smallest eigenvalue of ${\bf T}$ to become $\approx H$ at $z=750$.

 \subsection{Molecular data}
 \label{ss:moldata}

We require the bound-bound and bound-free Einstein coefficients $A_{ij}$ and $\alpha_i$.  These must be calculated since no tabulations of the full radiative dissociation spectrum from the B and C states is available (Ref.~\cite{2000A&AS..141..297A} gives integrated radiative dissociation rates and mean photon energies, but not a spectrum as required here).

For the energy levels and dipole transitions, we use the Born-Oppenheimer approximation.  This may not be accurate for nearly degenerate vibrational levels of the B and C$^+$ states: these can mix if they have the same rotational quantum number $J$ (C$^-$ states cannot mix with B by parity conservation).  In these cases, if e.g. the decay rate to a particular X rovibrational level is much greater for the pure B than the pure C$^+$ state, then the mixing can enhance the decay rate from the energy eigenstate that is ``mostly'' C$^+$ \cite{1989A&AS...79..313A, 2000A&AS..141..297A}.  However, in these cases we expect that the rates involving the B state would have a greater impact on the H$_2$ abundance than the C$^+$ state.  Therefore, the Born-Oppenheimer approximation is sufficient for the purpose of order-of-magnitude estimation of the H$_2$ abundance.

We have acquired the electronic energy level surfaces from the literature for the X \cite{1993JChPh..99.1851W}, B \cite{2002JMoSp.212..208S} and C \cite{2003JMoSp.220...45W} states, and used them to construct Born-Oppenheimer wave functions.  The electronic dipole matrix elements have also been obtained from the literature for X--B \cite{2003JMoSp.217..181W} and X--C \cite{2003JMoSp.220...45W} transitions.  (The expressions for Einstein coefficients can be found in Appendix~\ref{app:H2dipole}.)  Note that for the radiative dissociation of the B and C electronic states, we consider decays to the vibrational continuum of the X electronic state.

\begin{figure}
\includegraphics[width=3.2in]{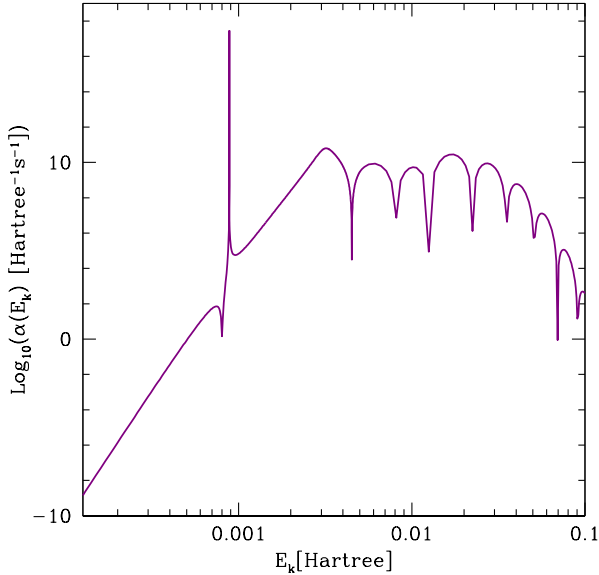}
\caption{\label{fig:AP_J14.nu9}The $P$-branch contribution to the spontaneous dissociation of H$_2$(B$^1\Sigma_u^+$) from the $J=14$, $\nu=9$ level, as a function of the center-of-mass frame kinetic energy of the final H atoms.  Note the quasibound resonance peak at $E_k=0.0009$ Hartree, which contains $\sim 4$\%\ of the integrated rate.}
\end{figure}

The dissociation coefficient $\alpha_i(E_f)$ is a complicated function of the final energy $E_f$ due to the existence of quasibound resonances of the H$_2$ X$^1\Sigma_g^+$ electronic state that ultimately dissociate via tunneling through the centrifugal barrier.  This behavior can still be captured by the Born-Oppenheimer approximation as long as one uses sufficiently small steps in $E_f$; see Fig.~\ref{fig:AP_J14.nu9}.  (In principle, there is an additional requirement that the resonances not be radiatively broadened, but since they have no allowed decays this is not a problem for us.)  In practice, when computing the integrals in Eq.~(\ref{equ:xdotbf}) we use the adaptive step size integrator {\tt odeint} of Ref. \cite{1992nrca.book.....P}.

We also consider the effect of adding the electric quadrupole transitions among different X$^1\Sigma_g^+$ states on the H$_2$ abundances by using the results of Ref.~\cite{1998ApJS..115..293W} for the corresponding Einstein coefficients.

 \subsection{Radiation field}
 \label{ss:rf}

The above equations require knowledge of the radiation field as a function of the photon energy, $f(E)$.  The relevant range of energies extends up to $0.54$~Hartree ($14.7$~eV).  This includes the range in which the spectral distortion is significant.  It also extends to energies greater than the ionization energy of H{\sc\,i}, $E_{\rm I}=13.6\,$eV$=$0.5 Hartree.

The CMB blackbody component is specified by the radiation temperature.  This is $T(z)=2.728 (1+z)$ Kelvin, and we have not distinguished between the matter temperature and the radiation temperature (in our redshift range $z>800$ these differ by $<0.1$\%). The number density of hydrogen nuclei is
\begin{equation}
n_{\rm H}(z)=\frac{(1-Y) \rho_{\rm cr} \Omega_{b,0}}{m_p} (1+z)^3,
\label{equ:nH}
\end{equation} 
where $Y=0.24$ is the mass weighted primordial helium abundance, $\rho_{\rm cr}=1.8788\times10^{-29} h^2$ g cm$^{-3}$ is the critical density of the universe at the present time, $m_p$ is the proton mass and $\Omega_{b,0} h^2=0.022$ is the baryon density parameter at the present time.

The radiation field at $E<E_{\rm I}$ is obtained using the code of Ref.~\cite{2008PhRvD..78b3001H}, which includes both CMB blackbody and spectral distortions.  This code self-consistently follows the absorption and emission of Lyman-series photons in the H{\sc\,i} lines.  In the case of two-photon transitions, it follows only the harder rather than the softer of the two photons.  This leads to small errors at $E<\frac38E_{\rm I}$, because the lower-energy photon in the decay
\begin{equation}
{\rm H(2s)} \leftrightarrow {\rm H(1s)} + \gamma + \gamma
\end{equation}
is ignored.  However, at the redshifts of interest here ($z>800$) this is not a significant oversight because the CMB blackbody is dominant at these low energies, and even at later times the distortion from redshifting of higher-energy photons (Lyman-$\alpha$ or the hard 2$\gamma$ photons) is more important than the soft 2$\gamma$ photons.  The code also includes the spectral distortion due to two-photon decays from higher levels H(3s,3d), and due to the Raman process
\begin{equation}
{\rm H(2s)} + \gamma \leftrightarrow {\rm H(1s)} + \gamma,
\end{equation}
which results in a significant addition to the photons at $E>E_{{\rm Ly}\alpha}$.

The code of Ref.~\cite{2008PhRvD..78b3001H} does not track the extreme ultraviolet (EUV), defined here as photons energetic enough to ionize hydrogen, $E\ge E_{\rm I}$.  The reason is that the Universe is optically thick to such photons: in $< 10^{-7}$ Hubble times these photons will be absorbed.  There is, however, a constant stream of EUV photons being produced via direct recombinations to the ground state,
\begin{equation}
{\rm H}^+ + e^- \leftrightarrow {\rm H(1s)} + \gamma_{\rm EUV}.
\end{equation}
The abundance of these photons will thus rapidly reach its equilibrium value \cite{2007A&A...475..109C}; in the limit of $f(E)\ll 1$ so that we can neglect stimulated recombinations, we have
\begin{equation}
f(E) = e^{-E/k_{\rm B}T} \frac{n[{\rm H}^+]n_e/n[{\rm H(1s)}]}{\{ n[{\rm H}^+]n_e/n[{\rm H(1s)}] \}_{\rm Saha}}.
\end{equation}
[The ``temperature'' in this equation (both in the exponential and the Saha abundance ratio) is technically the matter temperature $T_{\rm m}$ since the photons are being produced and destroyed by interaction with matter, but at high redshift we do not make this distinction.]

\begin{figure}
\includegraphics[width=3.2in]{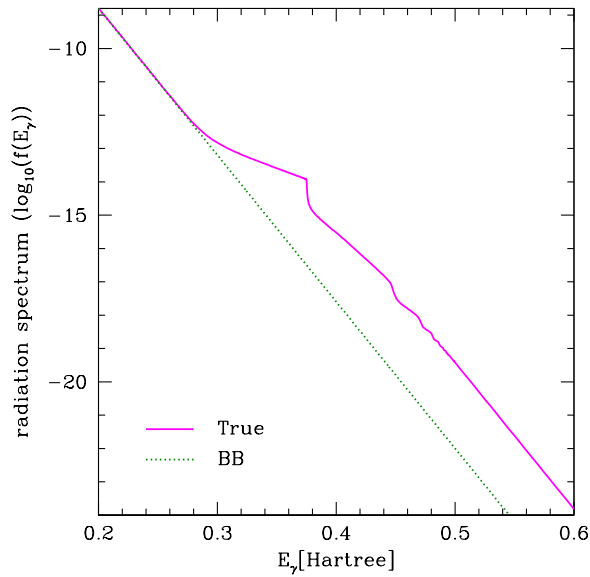}
\caption{\label{fig:fnuPlot}The radiation spectrum at $z=1142$, showing both the CMB blackbody spectrum (straight line) as well as the full calculation including H spectral distortions.}
\label{fig:fnu}
\end{figure}

The overall radiation spectrum is shown in Fig.~\ref{fig:fnu} at $z=1142$.

In our analysis, we take the output of the standard calculation, namely the abundance of hydrogen atoms and the phase space density of photons as a function of redshift and energy, as the input in our calculation of the abundance of hydrogen molecule levels. Of course, if H$_2$ abundances turn out to be high enough to make a considerable change in the recombination history or radiation spectrum, one must perform a more self-consistent calculation in which the feedback of H$_2$ molecules on hydrogen recombination is taken into account properly by solving the rate equation for all of the species simultaneously.  But for the standard cosmology, we will see that this situation does not arise.

\subsection{The charged-particle processes}
\label{ss:charged}

We have investigated the processes that produce H$_2$ from neutral hydrogen atoms and radiation.  However, there are other mechanisms that contribute to their formation and destruction, namely the H$^-$, H$_2^+$, and HeH$^+$ pathways (these have been found to be dominant at low $z$ in previous works, e.g. Refs.~\cite{1998A&A...335..403G, 2006MNRAS.372.1175H}). 

\begin{figure}
\includegraphics[width=3.2in]{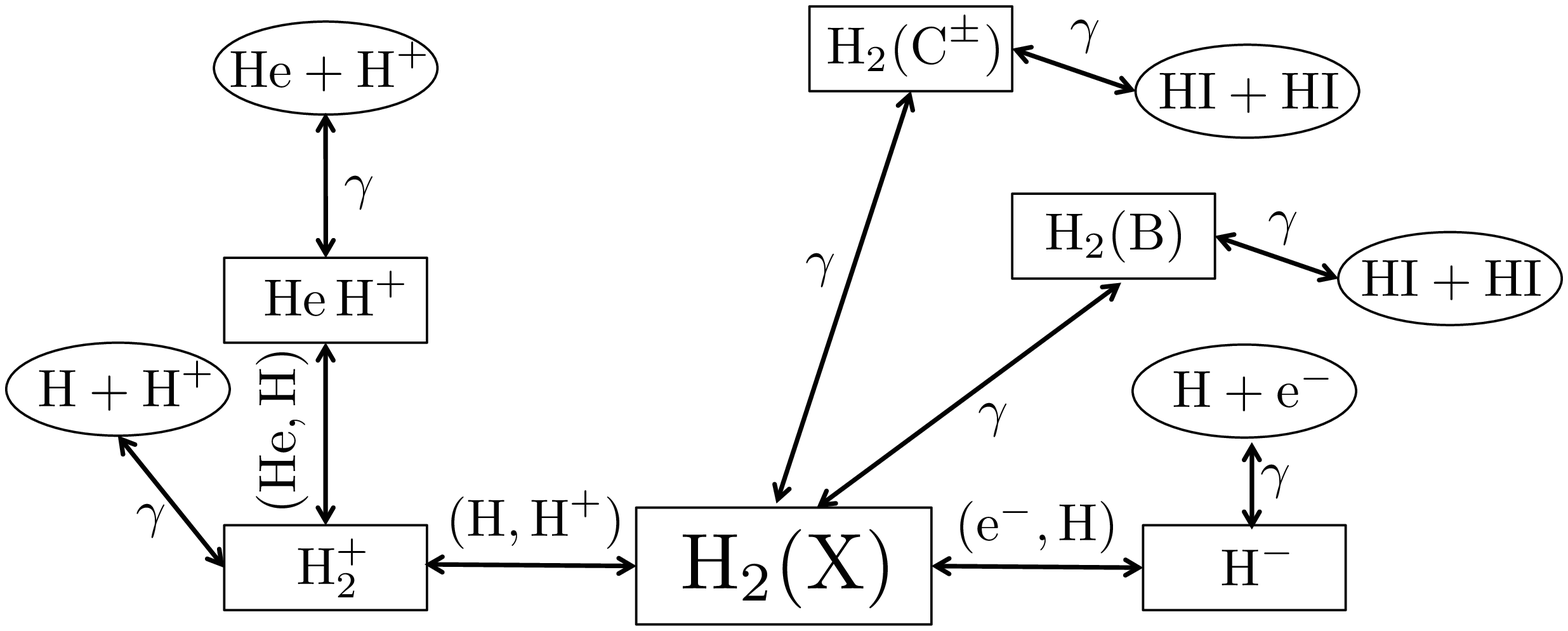}
\caption{The network of routes for the formation and destruction of an H$_2$ molecule that we consider in this paper. }
\label{fig:paths}
\end{figure}

All of the reactions that we consider in this paper, including these pathways, are shown in Fig.~\ref{fig:paths}. They include the radiative attachment/detachment for H$^-$

\begin{equation}
{\rm H} + e^- \leftrightarrow {\rm H}^- + \gamma,
\label{eq:R1}
\end{equation}
the radiative association/dissociation of H$_2^+$
\begin{equation}
{\rm H} + {\rm H}^+ \leftrightarrow {\rm H}^+_2 + \gamma,
\label{eq:R2}
\end{equation}
and that for HeH$^+$
\begin{equation}
{\rm He} + {\rm H}^+ \leftrightarrow {\rm HeH}^+ + \gamma.
\label{eq:R3}
\end{equation}

There are subsequent nonradiative reactions that generate H$_2$ (and at sufficient temperature can destroy it by operating in reverse): the H$^-$ channel
\begin{equation}
{\rm H}^- + {\rm H} \leftrightarrow {\rm H}_2 + e^-
\label{eq:R4}
\end{equation}
and the H$_2^+$ channel
\begin{equation}
{\rm H}_2^+ + {\rm H} \leftrightarrow {\rm H}_2 + {\rm H}^+.
\label{eq:R5}
\end{equation}
The latter can be aided by proton exchange from HeH$^+$:
\begin{equation}
{\rm HeH}^+ + {\rm H} \leftrightarrow {\rm He} + {\rm H}_2^+.
\label{eq:R6}
\end{equation}

\subsubsection{Positive channel: H$_2^+$ and HeH$^+$}

The positive ion channel (H$_2^+$) must be treated via level-resolved chemistry \cite{1998A&A...335..403G, 2002JPhB...35R..57L, 2006MNRAS.372.1175H}.  Our solution to this is to extend the ${\bf T}$-matrix to include the additional species.  That is, we write
\begin{equation}
{\bf T} = \left( \begin{array}{ccc} {\bf T}_{{\rm H}_2,{\rm H}_2} & {\bf T}_{{\rm H}_2,{\rm H}_2^+} & {\bf T}_{{\rm H}_2,{\rm HeH}^+} \\
{\bf T}_{{\rm H}_2^+,{\rm H}_2} & {\bf T}_{{\rm H}_2^+,{\rm H}_2^+} & {\bf T}_{{\rm H}_2^+,{\rm HeH}^+} \\
{\bf T}_{{\rm HeH}^+,{\rm H}_2} & {\bf T}_{{\rm HeH}^+,{\rm H}_2^+} & {\bf T}_{{\rm HeH}^+,{\rm HeH}^+}
\end{array} \right),
\end{equation}
and similarly the source vector ${\bf s}$ and abundance vector ${\bf x}$ are extended.  The model H$_2$ molecule includes 1435 levels (X, B, C$^+$, C$^-$) with $J\le 20$.  We follow all 423 rovibrational levels of H$_2^+$(X$^2\Sigma_g^+$).  However, we treat the HeH$^+$(X$^1\Sigma^+$) ion assuming Boltzmann distribution of the rovibrational levels at the radiation temperature, which is a good approximation since electric dipole transitions are allowed and rapidly thermalize the level populations.  The full ${\bf T}$-matrix is thus 1859$\times$1859.

The sub-block of the ${\bf T}$-matrix and ${\bf s}$-vector involving H$_2^+$ and HeH$^+$ was computed in Ref.~\cite{2006MNRAS.372.1175H}: it contains Eqs.~(\ref{eq:R2}), (\ref{eq:R3}), and (\ref{eq:R6}), as well as contributions associated with the electric quadrupole transitions among the rovibrational states of H$_2^+$(X$^2\Sigma_g^+$).  We used the code and rate coefficients of Ref.~\cite{2006MNRAS.372.1175H} to generate the corresponding matrices and source vectors.  Since none of our reactions directly connect HeH$^+$ to H$_2$ (they can only interconvert via H$_2^+$), we set ${\bf T}_{{\rm H}_2,{\rm HeH}^+} = {\bf 0}$ and ${\bf T}_{{\rm HeH}^+,{\rm H}_2} = {\bf 0}$.  It therefore remains only to determine the sub-blocks ${\bf T}_{{\rm H}_2,{\rm H}_2^+}$ and ${\bf T}_{{\rm H}_2^+,{\rm H}_2}$, which arise from Eq.~(\ref{eq:R5}).  These are related by the usual detailed balance relation,
\begin{equation}
\frac{R_{ji}}{R_{ij}} = \frac{x_{i,\rm th}}{x_{j,\rm th}} = \frac{x[{\rm H(1s)}]}{x[\rm{H}^+]}\frac{g_i}{2g_j} e^{(E_j+E_{\rm H(1s)}-E_i - E_{{\rm H}^+})/k_{\rm B}T},
\end{equation}
where $i$ represents any state of H$_2$, $j$ any state of H$_2^+$, $E_{\rm H(1s)}-E_{{\rm H}^+} = -13.6$ eV, and the factor of 2 comes from the degeneracy ratio of H versus H$^+$.  The corresponding contributions to ${\bf T}$ can then be determined from Eq.~(\ref{eq:Tdef}).

The rovibrational level-resolved forward reaction rates $R_{ij}$ for Eq.~(\ref{eq:R5}) are unfortunately not available in the literature.  Vibrationally resolved quantum-mechanical rates have been calculated \cite{2002PhRvA..66d2717K}, however their calculation did not resolve the rotational levels and did not well-sample the lowest energies required here.  There is also the experimentally measured low-temperature rate of $6.4\times 10^{-10}$ cm$^3$ s$^{-1}$ \cite{1979JChPh..70.2877K}, which once again did not resolve rovibrational levels (see also the recent measurements of the reaction of D$_2^+$+H \cite{2009JPhCS.194a2043A}).  The experiment of Ref.~\cite{1979JChPh..70.2877K} did however show by isotopic substitution that the reaction mechanism is charge transfer, i.e. the two nuclei in the initial H$_2^+$ ion remain in the H$_2$ molecule that is produced.  Therefore the ortho- or para- nuclear spin character should be preserved in the reaction of Eq.~(\ref{eq:R5}).

Since the calculations of Ref.~\cite{2002PhRvA..66d2717K} suggest that at energies of several tenths of an eV the reaction H$_2^+(\nu=0)$+H$\rightarrow $H$_2$+H$^+$ is most likely to leave the final molecule in the $\nu=4$ vibrational state, a simple prescription is to assume that (i) the rate coefficient for this reaction is $6.4\times 10^{-10}$ cm$^3$ s$^{-1}$; (ii) the final vibrational state is $\nu=4$; and (iii) the rotational quantum number ($N$ for H$_2^+$ and $J$ for H$_2$) is unchanged in the collision.

\subsubsection{Negative channel: H$^-$}

Previous work \cite{1998A&A...335..403G, 2002JPhB...35R..57L, 2006MNRAS.372.1175H} has established that at $z>200$, reaction Eq.~(\ref{eq:R1})  forces the H$^-$ abundance to its thermal equilibrium value because (i) the matter and radiation temperatures are equal, and (ii) photodetachment by blackbody photons is the main sink for H$^-$ on account of the strong CMB field and the low binding energy of H$^-$, $B[$H$^-]=0.754$ eV.  The thermal abundance is then given by
\begin{equation}
x[{\rm H}^-] = \frac{n_{\rm H}x_ex_{{\rm H}(1s)} }{4} \left( \frac{2\pi\hbar^2}{m_{\rm e}k_{\rm B}T} \right)^{3/2}
e^{B[{\rm H}^-]/k_{\rm B}T}.
\end{equation}

The associative detachment reaction, Eq.~(\ref{eq:R4}), can be incorporated by adding appropriate sources and sinks.  For the sources, we add
\begin{equation}
s_i += k_i n_{\rm H} x[{\rm H}^-] x[{\rm H}(1s)],
\label{eq:si+}
\end{equation}
where ``$+=$'' means that the quantity is added to $s_i$.  Here $k_i$ is the rate coefficient to level $i$ of H$_2$, which we obtain from the calculations of Ref.~\cite{1991A&A...252..842L}; we use the $T=3000\,$K column of their table as it is most appropriate for the recombination epoch and the temperature dependences are weak.  Detailed balance implies a corresponding sink for H$_2$ molecules:
\begin{equation}
T_{ii} += \frac{ k_i n_{\rm H} x[{\rm H}^-] x[{\rm H}(1s)] }{ x_{i,{\rm th}} }.
\label{eq:tii+}
\end{equation}
(We may use $x_{i,{\rm th}}$ here since the negative species are catalysts and hence have no effect if H$_2$ is at the thermal abundance; one may check explicitly that $T_{ii}$ in fact is proportional to $n_{\rm H(1s)}$ with a coefficient that depends only on temperature.)

\cmnt{
The production of H$_2$ then requires the rate coefficient for the associative detachment reaction Eq.~(\ref{eq:R4}), which we take as $1.5\times 10^{-9}(T/300$ K$)^{-0.1}$ cm$^3$ s$^{-1}$ \cite{1998ApJ...509....1S}.

The contribution of the positive ion channels (H$_2^+$, HeH$^+$) can be estimated using the code from Ref.~\cite{2006MNRAS.372.1175H}, which treats the above reaction network, fully taking account of all 423 rovibrational levels of H$_2^+$(X$^2\Sigma_g^+$) and the quadrupole transitions among these levels, but treating the HeH$^+$(X$^1\Sigma^+$) ion assuming Boltzmann distribution of the rovibrational levels at the radiation temperature (which is a good approximation since electric dipole transitions are allowed).  We used ``Model C'' of Ref.~\cite{2006MNRAS.372.1175H} for the level-resolved rate coefficients, but use of the alternatives (Models A or B) results in changes of at most $\sim 30$\%\ and hence would have no influence on our final conclusion that these species are not relevant to the recombination history.
}

 \section{results}
 \label{sec:R}

\subsection{H$_2$ abundance}

We now determine the H$_2$ abundance by two methods: first, with only the Lyman and Werner band reactions as sources and sinks for H$_2$; and second, including the charged particle reactions as well.  The latter, of course, is our final result.

\subsubsection{H$_2$ dipole and quadrupole transitions only}

The abundances of the H$_2$ molecule in different rovibrational states $x_i$ can be found by using Eqs.~(\ref{equ:beta}--\ref{equ:final}).

\begin{figure}
\includegraphics[width=3.2in]{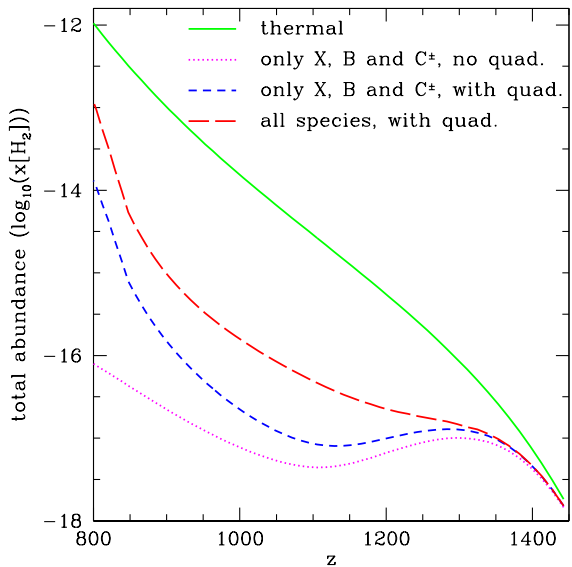}
\caption{The abundance of H$_2$ molecules, $x[$H$_2]$, as a function of redshift. Short dashed blue and dotted magenta lines show the cases where we do and do not include quadrupole transitions between X levels, respectively, and only the Lyman and Werner band reactions can create and destroy H$_2$. The long dashed red curve shows the case when all the relevant reactions are included. The green solid curve shows the thermal abundances of Eq.~(\ref{eq:thermalabundance}).}
\label{fig:sum_x}
\end{figure}

The total abundance of {\em all} H$_2$ molecules irrespective of rovibrational state is shown in Fig.~\ref{fig:sum_x}.  This abundance shows the expected rapid increase in the early stages of recombination as the temperature drops.  At $z\lesssim 1400$, the photon phase space density $f(E)$ begins to deviate substantially from the Planck spectrum.  One can see that the sense of the resulting spectral distortion is a temporary {\em decline} in the abundance of H$_2$, followed by a rapid recovery as the spectral distortion redshifts below the Lyman-Werner bands.

\begin{figure}
\begin{center}
\includegraphics[width=84mm]{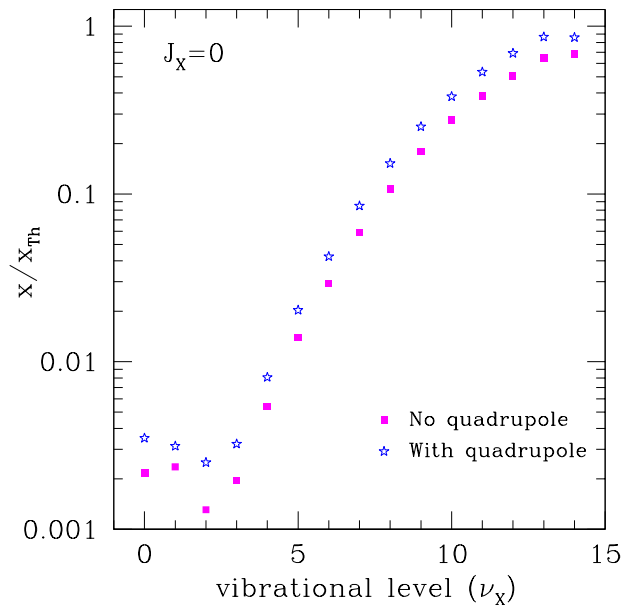}
\end{center}
\caption{The ratios of the abundances of the H$_2$ X$^1\Sigma_g^+$ levels with $J=0$ to their thermal values at $z=1142$, plotted as a function of the vibrational quantum number $\nu$.  Blue stars and magenta squares show the cases with and without quadrupole transitions between the X electronic levels, respectively.  This plot included only the Lyman and Werner band reactions as sources and sinks for H$_2$.}
\label{fig:ratio-all}
\end{figure}

We show in Fig. \ref{fig:ratio-all} the ratio $x_i/x_{i,\rm th}$ of the abundances of the states in all vibrational levels within the X$^1\Sigma_g^+$ $J=0$ sequence of the hydrogen molecule to their abundances in thermal equilibrium at $z=1142$.  Here again ``equilibrium'' refers to the abundance $x_{i,\rm th}$ that would be obtained if the reaction ${\rm H}_2({\rm X}^1\Sigma_g^+,\nu,J)\leftrightarrow 2{\rm H(1s)}$ were in equilibrium at the {\em actual} H(1s) abundance (the definition is important since the ionization fraction deviates from Saha).  The first two vibrational levels have approximately the same $x_i/x_{i,\rm th}$; then there is a sudden drop in this ratio going from the second to the third vibrational level, after which $x_i/x_{i,\rm th}$ increases steadily and approaches 1 for the weakly bound states.  The physical reason for this situation is that a photon in or redward of the H{\sc\,i} Lyman-$\alpha$ line is capable of exciting an H$_2$ molecule only from the $\nu\ge 2$ vibrational levels (starting from $\nu=0,1$ there is insufficient energy to reach the B or C electronic states).  Thus the $\nu\ge 2$ vibrational levels are rapidly photodissociated.  The inclusion of the quadrupole transitions enables H$_2$(X$^1\Sigma_g^+$) molecules to switch among the various vibrational levels and leads to a washing-out of the step in abundance versus $\nu$.

\subsubsection{Inclusion of charged particle reactions: H$_2^+$, HeH$^+$, and H$^-$}

\begin{figure*}
\plottwo{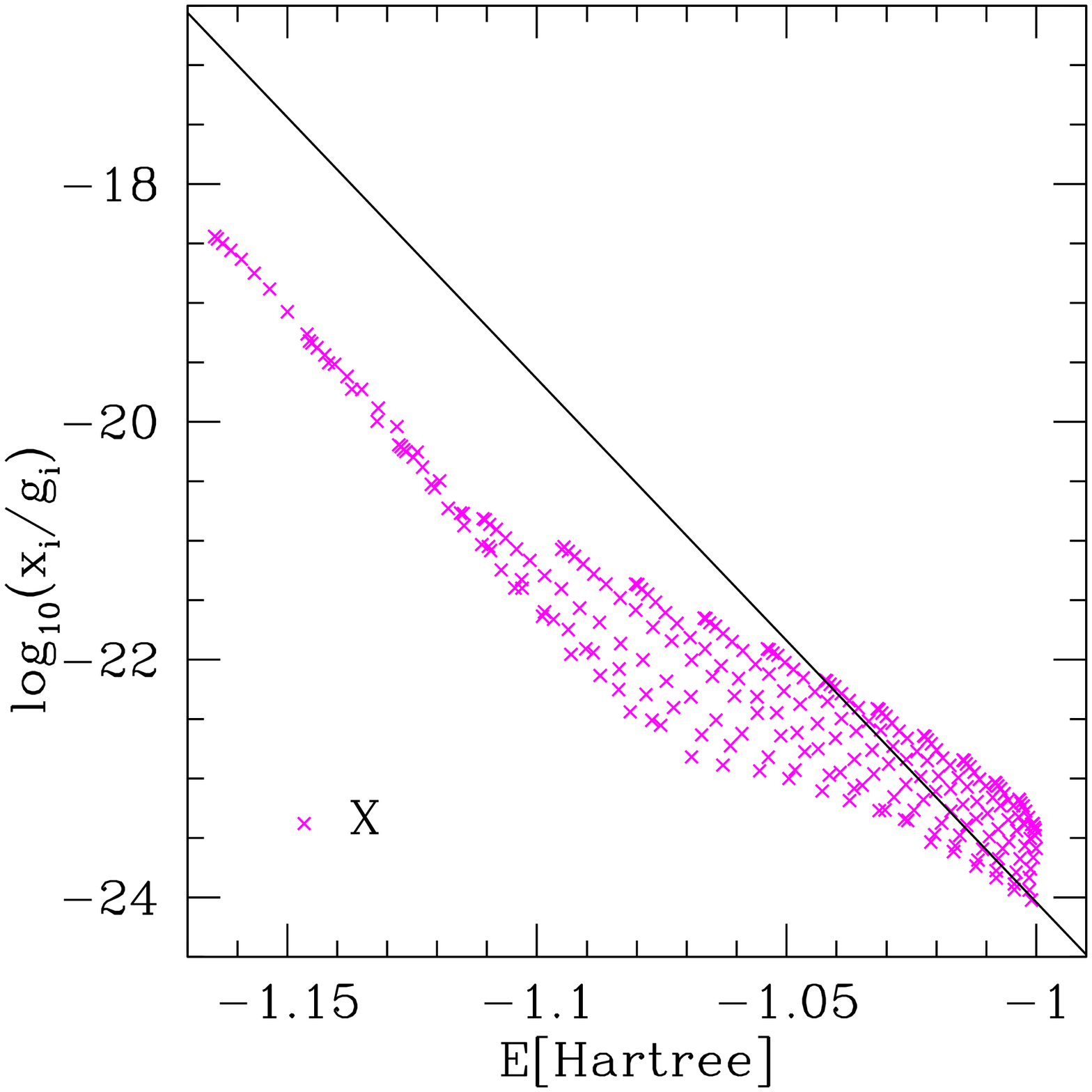}{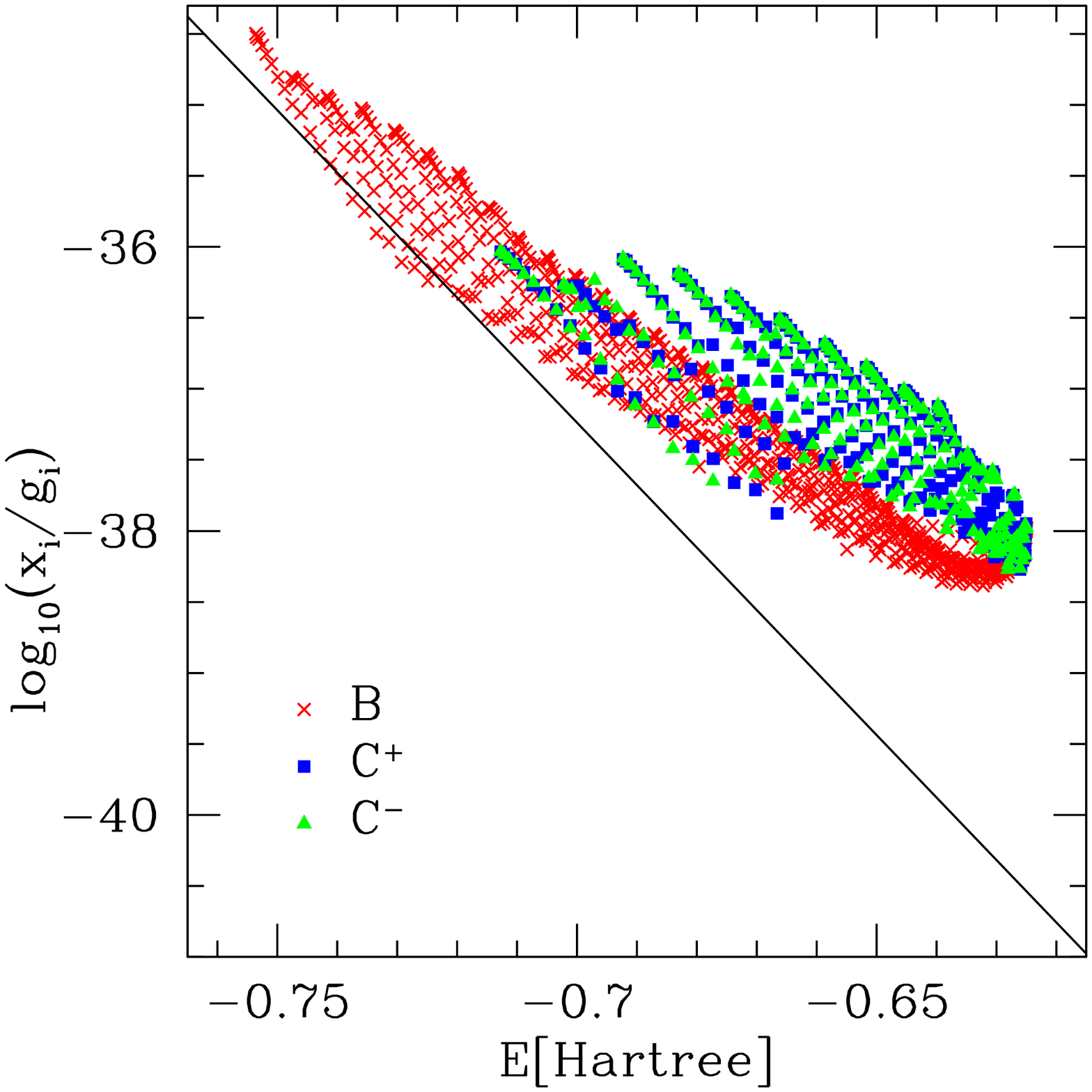}
\caption{\label{fig:scatter}The abundance of each H$_2$ level plotted versus the level energy at $z=1142$. In the left panel we show the abundances of X levels and in the right panel the B, C$^+$ and C$^-$ levels. The solid line shows the thermal equilibrium abundances.}
\end{figure*}

We now turn on our full reaction network. We display the abundances of the various levels of H$_2$ by plotting the logarithmic abundance $\log_{10}(x_i/g_i)$ versus the level energy.  This is a straight line in the case of a thermal distribution of levels.  The actual result is shown in Fig.~\ref{fig:scatter} for the X electronic states on the left and the B, C$^+$ and C$^-$ on the right. We can see that the highly excited rovibrational levels of the X electronic state are near thermal equilibrium, but the lower levels are underpopulated by $\sim 2$ orders of magnitude. On the other hand the B, C$^+$ and C$^-$ states are overpopulated compared to the equilibrium abundance and they become more overpopulated relative to thermal for higher energy levels.  This is a consequence of the shape of the spectral distortion.

\begin{figure}
\includegraphics[width=3.2in]{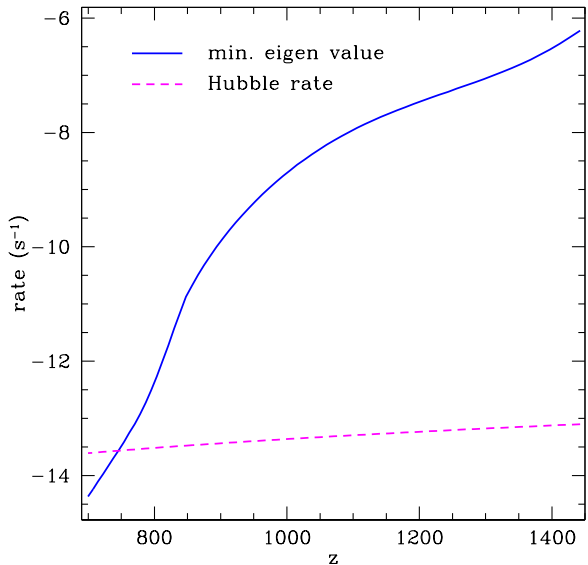}
\caption{\label{fig:EV}The minimum eigenvalue of the rate matrix $T$ (solid blue) as compared to the Hubble rate (dashed magenta).}
\end{figure}

Thus far we have assumed the steady state approximation, i.e. that all of the eigenvalues of the ${\bf T}$-matrix are large compared to the Hubble expansion rate $H$. The minimum eigenvalue of ${\bf T}$ is shown in Fig.~\ref{fig:EV} together with the Hubble rate. We can see that for $z>800$ the minimum eigenvalue $\lambda_{\rm min}$ is much larger than the Hubble rate.  In fact, due to the rapid increase in the H$_2$ abundance, we should really be comparing $\lambda_{\rm min}$ to $d\ln x[$H$_2]/dt$; the crossing of these occurs at $z\approx 810$, which is roughly where we expect the steady state approximation to fail.  As a consequence, we do not show H$_2$ abundances at lower redshifts.
 
In Fig.~\ref{fig:sum_x} we show the sum of the abundances of all the X, B and C$^{\pm}$ levels of the H$_2$ molecule as a function of redshift (in practice, this sum is dominated by the X electronic state). Short dashed blue and dotted magenta lines show the cases where we do and do not include quadrupole transitions between X levels, respectively, and only the Lyman and Werner band reactions can create and destroy H$_2$. The long dashed red curve shows the case when all the relevant reactions are included. The green solid curve shows the thermal abundances of Eq.~(\ref{eq:thermalabundance}). We can see that the addition of new transitions consistently causes the non-thermal abundance of H$_2$ molecules to increase, while always staying below the thermal abundance.

\begin{figure*}
\plottwo{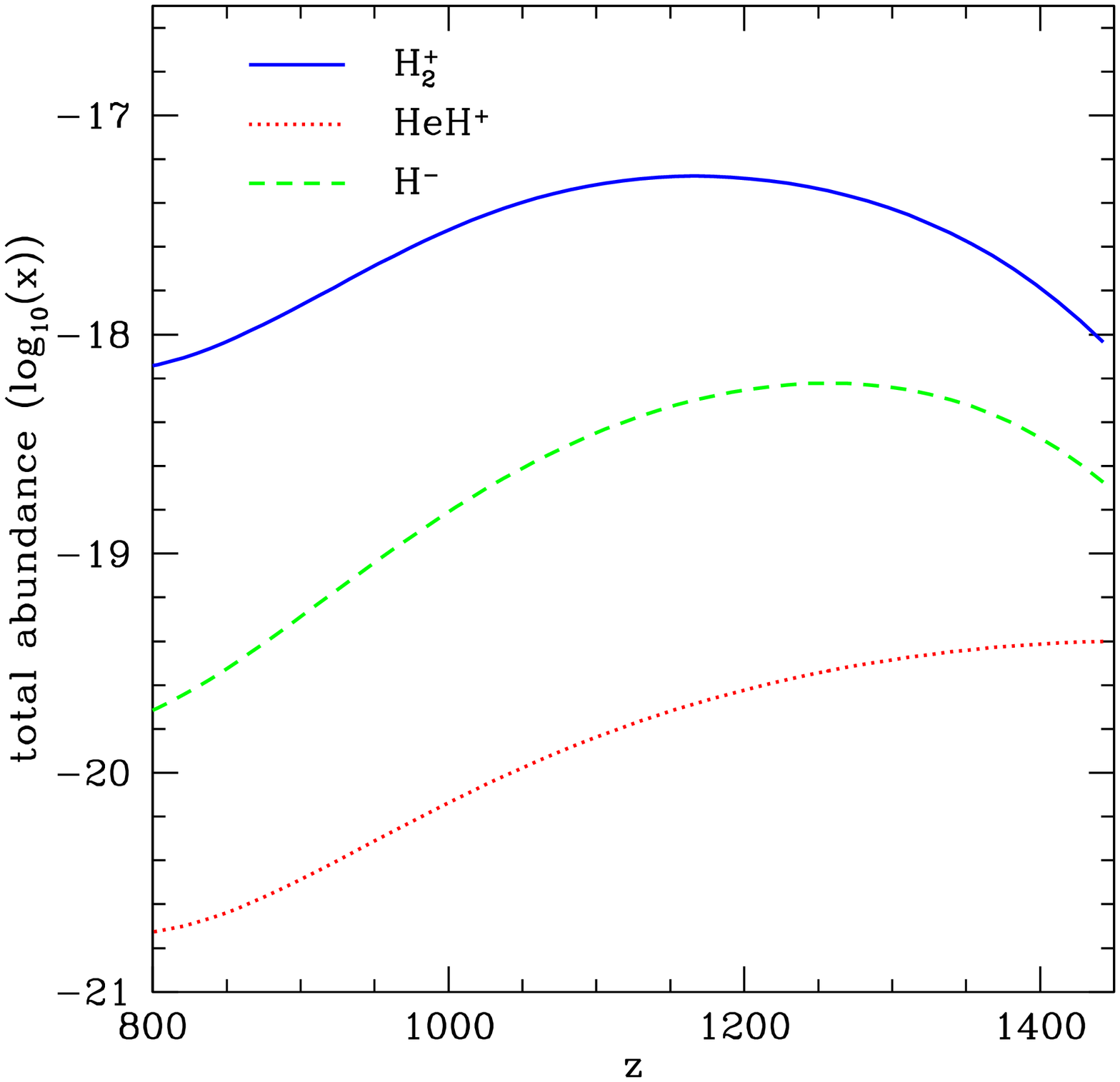}{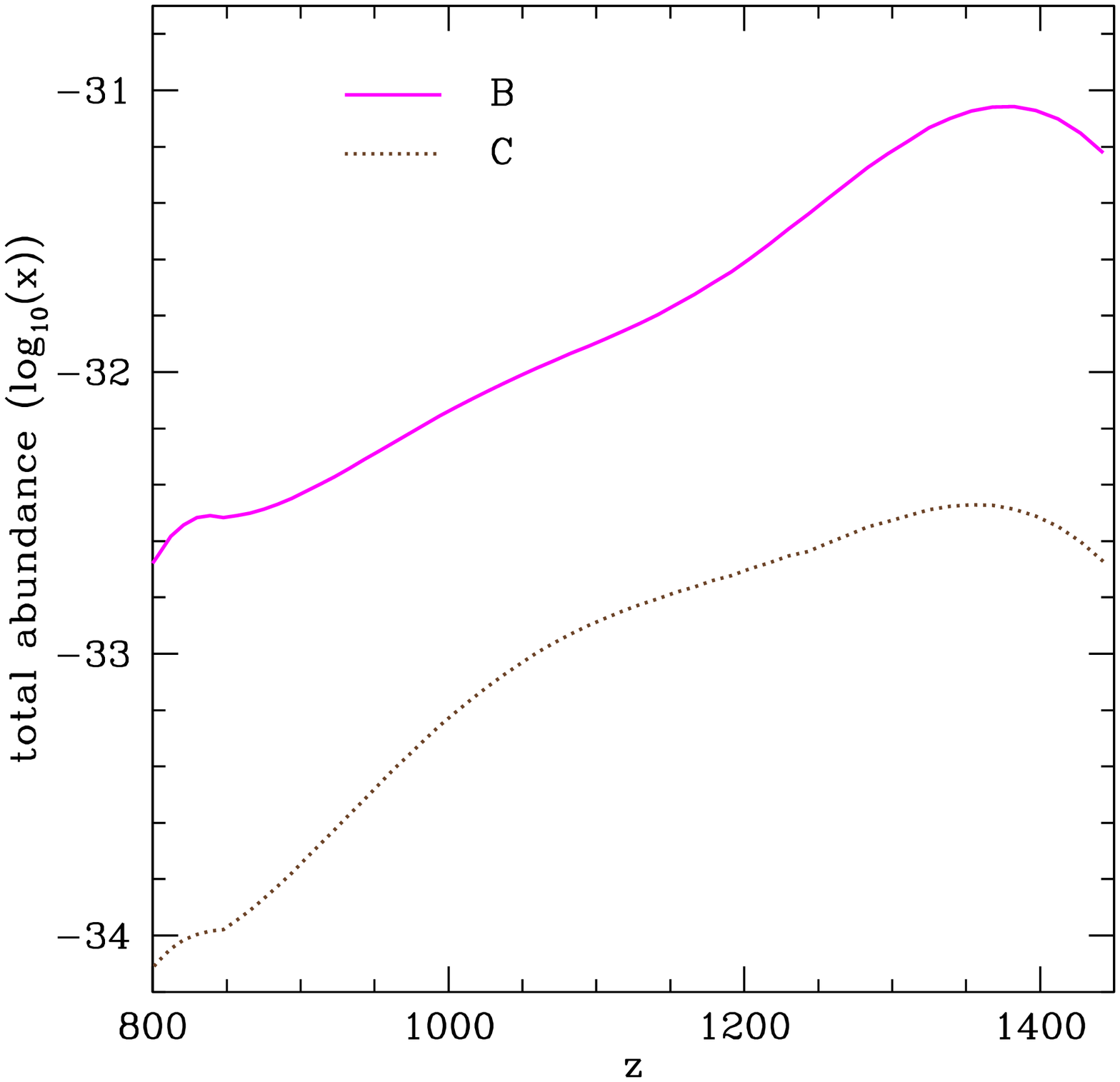}
\caption{\label{fig:intermed}Here we show the abundances of the intermediate species, H$^-$, H$^+_2$, and HeH$^+$ in the left panel and H$_2$(B,C$^{\pm}$) in the right panel, as a function of redshift during the recombination epoch.}
\end{figure*}

We also show the abundances of the intermediate species in Fig.~\ref{fig:intermed}. The left panel shows the abundances of the charged particles and the right panel the excited electronic states B and C$^\pm$ of the hydrogen molecule. Even though the abundances of the excited states are much smaller than those of the charged particles, they must be included since the transition rates between these excited states and the H$_2$ ground state can be much higher than the transition rates connecting the charged particles to the ground state. 

A related possible formation channel for H$_2$ that is mentioned in the literature is from the reaction of an excited and a ground-state H atom \cite{2006ApJ...646L..91D}:
\begin{equation}
{\rm H}(1s) + {\rm H}(n=2) \leftrightarrow {\rm H}_2({\rm X}^1\Sigma_g^+) + \gamma.
\end{equation}
This reaction proceeds if the reactants approach each other in the vibrational continuum of a $^1\Sigma_u^+$ or $^1\Pi_u$ electronic state (usually B or C), and in order to produce a bound H$_2$ molecule the photon must have an energy $E>E($Ly$\alpha)$.  Since the phase space density of photons is a steeply decreasing function of energy (including a step at Lyman-$\alpha$), we would expect that the photodissociation of the levels of H$_2$ is dominated by transitions through the discrete vibrational levels of B$^1\Sigma_u^+$ and C$^1\Pi_u$ rather than the higher-energy continuum.  One can also check the formation of H$_2$ by this mechanism directly: using the rate coefficient $2.09\times 10^{-14}(T/300$ K$)^{0.24}e^{-T/37800\,{\rm K}}$ cm$^3$ s$^{-1}$ \cite{1998ApJ...509....1S}, we find a maximum production rate of H$_2$ molecules per H nucleus per Hubble time of $7\times 10^{-12}$ at $z=1250$.  This channel is thus subdominant to the HeH$^+$/H$_2^+$ channel.  All of these are small compared to the transitions through H$_2$(B$^1\Sigma_u^+$,C$^1\Pi_u$).

\subsection{Effect on recombination}

Since some of the X--B and X--C transitions have energy above the Lyman-$\alpha$ energy, the absorption or emission of a photon in this line by an H$_2$ molecule adds or removes a photon that would have otherwise excited a hydrogen atom at a later time, and therefore it will alter the process of recombination of hydrogen atoms.  To estimate this effect, we must estimate both the Sobolev optical depth of the various lines using Eq.~(\ref{equ:Sobolev}), and the rate of production of photons in the X--B and X--C bands.

\begin{figure}
\includegraphics[width=3.2in]{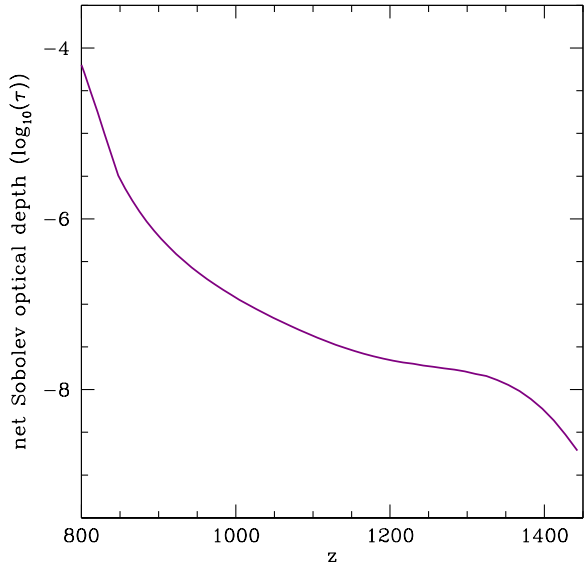}
\caption{\label{fig:tau}Total Sobolev optical depth of all of the Lyman and Werner lines as a function of redshift.}
\end{figure}

In Fig.~\ref{fig:tau} we show the net Sobolev optical depth for all X--B and X--C lines. We can see that the optical depth of the sum of all the lines is much smaller than one, and therefore each individual line is also optically thin. This then justifies using $P_{ij}=1$ for the Sobolev escape probability throughout this paper.

\begin{figure}
\includegraphics[width=3.2in]{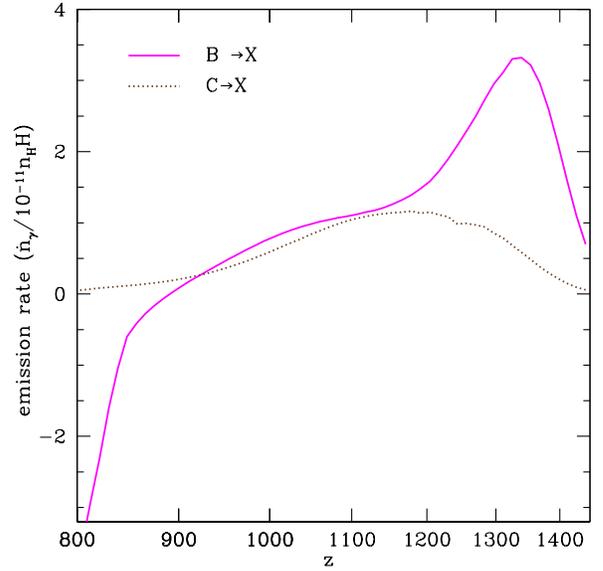}
\caption{\label{fig:emission}The net rate of emission (i.e. emission minus absorption) of photons at $E>E($Ly$\alpha)$ in the H$_2$ Lyman and Werner bands, measured in photons per H nucleus per Hubble time.}
\end{figure}

The net rate of emission of photons in the X--B and X--C bands at energies above the Lyman-$\alpha$ energy is shown in Fig.~\ref{fig:emission}.  We see that there was no absorption or emission in this line at very early times when the universe was in thermal equilibrium. However, at redshifts of $\sim 1500$ the deviation of the hydrogen molecule abundances from their thermal values, coupled with the CMB spectral distortion, lead to a net emission of super Ly$\alpha$ photons for both the Lyman and Werner bands. The Lyman band, however, started absorbing super Ly$\alpha$ radiation later on. The net number of emitted photons turns out to be negligible ($\sim 10^{-11}$) and so it can only cause a very small change in the abundance of H atoms.

There is one final possibility for reactions involving H$^-$ and H$_2^+$ to affect recombination, namely through the neutralization reactions
\begin{equation}
{\rm H}^- + {\rm H}^+ \leftrightarrow 2{\rm H}
\label{eq:R7}
\end{equation}
and
\begin{equation}
{\rm H}_2^+ + e^- \leftrightarrow 2{\rm H},
\label{eq:R8}
\end{equation}
which -- in combination with Eqs.~(\ref{eq:R1}) and (\ref{eq:R2}) -- lead to a net recombination.  They are, however, negligible: even for the ``upper limit'' rate coefficient for Eq.~(\ref{eq:R7}) of $5\times 10^{-9}$ cm$^3$ s$^{-1}$ \cite{2006ApJ...640..553G}, we obtain a maximum recombination rate per Hubble time ($\dot x_{1s}/H$) of $1.1\times 10^{-11}$.  Even this is too large since Eq.~(\ref{eq:R7}) usually leaves one of the hydrogen atoms in an excited level (principally $n=3$) \cite{1986JPhB...19L..31F} from which it has a large probability of being photoionized.  Similarly, for Eq.~(\ref{eq:R8}), the rate coefficient at recombination-epoch temperatures is of order $\sim 10^{-8}$ cm$^3$ s$^{-1}$ \cite{1994ApJ...424..983S}, which implies recombination rates $\dot x_{1s}/H$ of order $10^{-10}$.
 
 \section{Discussion and Conclusion}
 \label{sec:DandC}

Since the Lyman and Werner band transition energies of H$_2$ are near the H Ly$\alpha$ energy, it is expected that the abundances of H$_2$ energy levels deviate appreciably from their thermal abundances.  This is because the photon phase space density has been distorted by the redshifted Ly$\alpha$ photons as in the standard hydrogen atom recombination picture.  However, it is not clear from the outset whether this distortion to the photon phase space density increases or decreases the abundances of H$_2$ levels compared to their thermal abundances, since the spectral distortion photons accelerate both the production and destruction of H$_2$.  To answer this question and ultimately to see to what extent the H$_2$ molecules can affect the recombination history we have in this paper carried out a detailed calculation including all of the rovibrational levels of the H$_2$ X$^1\Sigma_g^+$, B$^1\Sigma_u^+$, and C$^1\Pi_u$ electronic states up to rotational number $J=20$, together with the charged species relevant to the formation of hydrogen molecules, that is H$_2^+$, HeH$^+$ and H$^-$. We have calculated the bound-bound and bound-free dipole transition rates for the Lyman and Werner bands of the hydrogen molecule using the Born-Oppenheimer approximation. Special care has been taken to find the resonances of the bound-free transitions. The rate equations connecting the energy levels are then solved in the steady state approximation and the level abundances are found by a matrix inversion for each given redshift.

The main result of our paper is that the shape of the CMB spectral distortion reduces the abundance of H$_2$ compared to the thermal abundance, resulting in low H$_2$ abundances throughout the recombination epoch; see Fig.~\ref{fig:sum_x}. The inclusion of the quadrupole transitions among rovibrational levels of the X electronic state increases the H$_2$ abundance, and adding the charged particle processes increases the H$_2$ abundances yet more, while remaining below the thermal abundance. We find $x[$H$_2]\sim 10^{-16}$ during most of the recombination epoch, rising to $10^{-13}$ at $z=800$.  We conclude that -- despite the high cross section for Lyman and Werner band absorption -- H$_2$ is not relevant for determination of the primordial recombination history and CMB anisotropies.
 
 \section*{Acknowledgements}
 
 E.A. thanks Ben Wandelt for his support, Laura Book for her constant help,  and Yacine Ali-Ha\"{\i}moud and Dan Grin for useful conversations. He acknowledges financial support from the National Science Foundation (grant AST 07-08849). C.H. is supported by the U.S. Department of Energy (DOE-FG03-92-ER40701), the National Science Foundation (NSF AST-0807337), and the David \& Lucile Packard Foundation.
 
\appendix

\section{Formulae for H$_2$ dipole transitions}
\label{app:H2dipole}

This appendix summarizes the notation and formulae we use for the H$_2$ molecular wave function and the dipole transition rates.

We use unprimed coordinates $(x,y,z)$ to denote a laboratory-fixed frame, and primed $(x',y',z')$ to denote a coordinate system that rotates with the molecule: the $z'$-axis is taken to be parallel to the internuclear separation vector ${\bf R}$, and the $x'$ axis is then chosen to lie in the $zz'$-plane.

For the X$^1\Sigma_g^+$ electronic state, the wavefunctions can be written in the Born-Oppenheimer approximation as
\begin{equation}
\Psi_{X,\nu JM}(\mathbf{r},\mathbf{R})= \frac{\phi_{\nu J}(R)}{R} Y_{JM}(\hat{\bf R}) \chi({\bf r}|{\bf R}),
\end{equation}
where ${\bf r}=({\bf r}_1,{\bf r}_2)$ is the 6-dimensional vector of positions of the two electrons, and
$\chi({\bf r}|{\bf R})$ is the electronic energy eigenstate associated with the X$^1\Sigma_g^+$ state with internuclear separation ${\bf R}$.  A similar equation holds for the B$^1\Sigma_g^+$ electronic state.  For C$^1\Pi_u$, one must account for the existence of two degenerate wavefunctions $\chi_\pm({\bf r}|{\bf R})$ with $z'$-component of electronic orbital angular momentum $\Lambda=\pm1$.  In this case the wave function is
\begin{eqnarray}
\Psi_{C^{\pm},\nu JM}(\mathbf{r},\mathbf{R}) \!\!\! &=& \!\!\! \frac{\phi_{\nu J}(R)}{\sqrt{2} R} 
\Bigl[ Y^{+1}_{JM}(\hat{\mathbf{R}}) \chi_{+}(\mathbf{r}|\mathbf{R}) 
\nonumber \\
&& 
\pm Y^{-1}_{JM}(\hat{\mathbf{R}}) \chi_{-}(\mathbf{r}|\mathbf{R}) \Bigr],
\end{eqnarray}
where $Y^s_{JM}$ is a spin-$s$ spherical harmonic \cite{1966JMP.....7..863N, 1967JMP.....8.2155G}.  (Some references, e.g. Ref.~\cite{1989A&AS...79..313A}, use rotation matrices instead of spin-weighted spherical harmonics, but these are equivalent.)  Note that there are two such wave functions ($+$ and $-$) with opposite parity, which are degenerate ($\Lambda$-doubled) in our level of approximation.

Normalization of the total wavefunction then requires $\int_0^{\infty} |\phi_{\nu J}(R)|^2 dR =1 $. Since the C electronic states have parity u, i.e. $\chi(-\mathbf{r}|\mathbf{R})=-\chi(\mathbf{r}|\mathbf{R})$, the total wavefunction satisfies: $\Psi_{C^{\pm},\nu JM}(\mathbf{r},-\mathbf{R})=\pm(-1)^{J+1} \Psi_{C^{\pm},\nu JM}(\mathbf{r},\mathbf{R})$.  

The Einstein coefficients for dipole transitions are discussed in general textbooks, e.g. Refs.~\cite{1965qume.book.....L, 1980MINTF...4.....B}.  In our case, the C$\rightarrow$X transition rates are as follows: for the electric dipole transition of a H$_2$ molecule from an excited state $a=(C,\nu_a,J_a)$ to a lower level $b=(X,\nu_b,J_b)$, we have
\begin{equation}
A_{ab}= \frac{4 e^2 (E_{a}-E_{b})^3}{3 \hbar^4 c^3 (2 J_a +1)} \left|\mathcal{M}_{ab}\right|^2,
\end{equation}
where the matrix element is
\begin{equation}
\left|\mathcal{M}_{ab}\right|^2=
\sum_{M_a=-J_a}^{J_a} \sum_{M_b=-J_b}^{J_b}
 \left| \left\langle \Psi_{X\nu_b J_b M_b} \left| \mathbf{d} \right| \Psi_{C \nu_a J_a M_a}\right\rangle \right|^2.
\end{equation}
Here ${\bf d} = -e{\bf r}_1-e{\bf r}_2$ is the electric dipole operator.  The matrix element expands as
\begin{eqnarray}
&&\!\!\!\!\!\!\!\! \!\!\!\! \left\langle \Psi_{X\nu_b J_b M_b} \left| \mathbf{d} \right| \Psi_{C^{\pm} \nu_a J_a M_a}\right\rangle
\nonumber \\
\!\!\! &=& \!\!\! \int d^3\mathbf{R} \frac{\phi^\ast_{X\nu_b J_b}(R)}{R} \frac{\phi_{C\nu_a J_a}(R)}{\sqrt{2}R} Y_{J_b M_b}^\ast(\hat{\mathbf{R}}) \nonumber \\
&& \!\!\! \times \left[ Y^{+1}_{J_a M_a} (\hat{\mathbf{R}}) \mathbf{D}^{CX}_{+}(\mathbf{R}) \pm Y^{-1}_{J_a M_a}(\hat{\mathbf{R}}) \mathbf{D}^{CX}_{-}(\mathbf{R})  \right],\;\;\;\;\;\;
 \label{equ:M}
\end{eqnarray}
where the fixed-${\bf R}$ electric dipole moment for the $C \rightarrow X$ transition is:
\begin{eqnarray}
\mathbf{D}^{CX}_{\pm}(\mathbf{R})
\!\!\! &=& \!\!\!
-e \int d\mathbf{r}^3_1 d\mathbf{r}^3_2 \chi^\ast_{X}(\mathbf{r}|\mathbf{R}) (\mathbf{r_1}+\mathbf{r_2}) \chi_{C,\pm}(\mathbf{r}|\mathbf{R}) \nonumber \\
&=& \!\!\! D_{CX}(R) \frac{\hat{\bf x}' \pm \hat{\bf y}'}{\sqrt{2}}.
\label{equ:D}
\end{eqnarray}
The values of $D_{CX}(R)$ are calculated in \cite{2003JMoSp.220...45W}. For the B$\rightarrow$X transition one finds $\mathbf{D}^{BX}(\mathbf{R})=D_{BX}(R) \hat{\mathbf{R}}$. We use the results of \cite{2003JMoSp.217..181W} for it.

By plugging Eq.~(\ref{equ:D}) into Eq.~(\ref{equ:M}), separating the $R$ integration from the $\hat{\mathbf{R}}$ integration, writing $\hat{\bf x}'$ and $\hat{\bf y}'$ in terms of the fixed coordinated unit vectors ($\hat{\bf x}$,$\hat{\bf y}$,$\hat{\bf z}$) with coefficient written in the form of spin wighted spherical harmonics with degree $s=0$, calculating the integrals of the three spin-weighted spherical harmonics in terms of the Wigner $3j$ symbols and using the orthogonality relations of the $3j$ symbols to do the sums over $M_a$ and $M_b$ one finally finds:
\begin{eqnarray}
|\mathcal{M}^P_{ab}|^2 &=& J_a K(C^+\nu_a J_a,X\nu_b J_b) , \nonumber \\
|\mathcal{M}^R_{ab}|^2 &=& (J_a+1) K(C^+\nu_a J_a,X\nu_b J_b)  , {\rm ~~and} \nonumber \\
|\mathcal{M}^Q_{ab}|^2 &=& (2J_a+1) K(C^-\nu_a J_a,X\nu_b J_b) ,
\end{eqnarray} 
where the change in angular momentum is denoted by the branch indices $P$ ($J_b-J_a=1$), $Q$ ($J_b-J_a=0$), and $R$ ($J_b-J_a=-1$), and
\begin{eqnarray}
\!\!\! && \!\!\!\!\!\!\!\!\!\!\!\!\!\!\!
K(Y_a \nu_a J_a, Y_b\nu_b J_b)
\nonumber \\
&=& \!\!\! \left|\int_0^{\infty} \phi_{Y_b \nu_b J_b}^\ast(R) D_{AB}(R) \phi_{Y_a \nu_a J_a}(R) dR \right|^2.
\end{eqnarray}

Similarly, for the B$\rightarrow$X transitions,
\begin{eqnarray}
|\mathcal{M}^P_{ab}|^2&=& (J_a+1) K(B\nu_a J_a,X\nu_b J_b)  {\rm ~~and}\nonumber \\
|\mathcal{M}^R_{ab}|^2&=& J_a K(B\nu_a J_a,X\nu_b J_b) .
\end{eqnarray}
These equations agree with Ref.~\cite{1989A&AS...79..313A}, appropriately restricted to the case of no B--C mixing.

\bibliographystyle{h-physrev}
\bibliography{ref}

\end{document}